# Development and characteristics of the HANARO ex-core neutron irradiation facility for applications in the boron neutron capture therapy field

**Myong-Seop Kim, Byung-Chul Lee, Sung-Yul Hwang, Heonil Kim and Byung-Jin Jun**

Korea Atomic Energy Research Institute, 150 Deokjin-Dong, Yuseong, Daejeon, 305-353, KOREA

E-mail: mskim@kaeri.re.kr

**Abstract.** The HANARO ex-core neutron irradiation facility was developed for various applications in the boron neutron capture therapy (BNCT) field, and its characteristics have been investigated. In order to obtain a sufficient thermal neutron flux with a low level contamination of fast neutrons and gamma-rays, a radiation filtering method is adopted. The radiation filter has been designed by using a silicon single crystal cooled by liquid nitrogen and a bismuth crystal. The installation of the main components of the irradiation facility and the irradiation room are finished. Experimental measurements of the neutron beam characteristics have been performed by using bare and cadmium covered gold foils and wires. The in-phantom neutron flux distribution was measured for a flux mapping inside the phantom. The gamma-ray dose was determined by using TLD-700 thermoluminescence dosimeters. The thermal and fast neutron fluxes and the gamma-ray dose were calculated by using the MCNP code, and they were compared with experimental data. The thermal neutron flux and Cd ratio which can be obtained at this facility are $1.49 \times 10^9$ n/cm$^2$s and 152, respectively. The maximum neutron flux inside the phantom was measured to be $2.79 \times 10^9$ n/cm$^2$s at a depth of 3 mm in the phantom. The two-dimensional in-phantom neutron flux distribution was determined, and the significant neutron irradiation was observed within 20 mm from the phantom surface. The gamma-ray dose rate for the free beam condition is expected to be about 80 cGy/hr. These experimental results are reasonably well supported by the calculated values of the facility design code. This HANARO thermal neutron facility can be used not only for a clinical trial but also for various basic irradiation researches of the BNCT field.

## 1. Introduction

Boron neutron capture therapy (BNCT) is a promising method for a cancer treatment in principle, which kills the cancer cells selectively by the use of a cancer seeking boron compound and a neutron irradiation (Slatkin 1991). This therapy is expected to be very effective for several types of cancers such as a malignant brain tumor-glioblastoma multiforme (GBM), for which no successful treatment has been developed. Now, worldwide research by using a research reactor and an accelerator is on going (Sauerwein *et al* 2002, Zamenhof *et al* 2004).

A cancer treatment by using BNCT is based on a reaction of the B-10 nucleus in a cancer cell with the neutron. Since the cross-section for the $B^{10}(n,\alpha)Li^7$ reaction almost decreases to $E^{-1/2}$, where E is the neutron energy, the discriminated dose between the normal and the cancer cells is mainly caused by the thermal neutron. However, the thermal neutron cannot penetrate deep into the tissue while the GBM has usually invaded deep into the normal brain tissue. Therefore, a neutron beam with a higher energy can be more effective than a direct irradiation of thermal neutrons. Moreover, an irradiation of these neutrons causes a thermal neutron peak at a depth of several cm. Since the biological damage of a neutron for a normal tissue is almost constant up to 10 keV, but rises rapidly above this level due to the dose from a proton recoil, the use of epithermal neutrons with an energy of a few keV is supposed to be the most effective for the BNCT of deep-seated tumours (Bisceglie *et al* 2000). Therefore, a whole brain irradiation by using epithermal neutrons without a debulking has been studied. On the other hand, a local irradiation with thermal neutrons after a debulking has been maintained with several methods to overcome the poor penetrability of thermal neutrons including a mixed irradiation of thermal and epithermal neutron beams (Sakurai *et al* 2000, Yamamoto *et al* 2002).

In HANARO, a 30 MW multipurpose research reactor, due to its design characteristics, a sufficient amount of epithermal neutrons for the BNCT cannot be obtained, but a sufficient amount of thermal neutrons can be obtained by means of a fast neutron and gamma-ray filtering method. We have designed an ex-core thermal neutron irradiation facility for various applications including BNCT. Over the past few years, the installation of the irradiation facility has been completed, and the irradiation room has been constructed.

The basic facilities required in a research reactor for a BNCT are a neutron irradiation facility for a treatment and a device for an instant measurement of the boron concentration in patients' blood samples. For HANARO, the ex-core thermal neutron irradiation facility is ready for a BNCT irradiation. And also, the facility for measuring the boron concentration by using a prompt gamma activation analysis with a good sensitivity has been developed (Byun *et al* 2002). This thermal neutron field can be used not only for a BNCT but also for the several neutron applications such as dynamic neutron radiography. For the BNCT applications, this facility can be used for a local irradiation and for various researches before clinical trials such as a small animal irradiation.

In this study, the design characteristics of the ex-core neutron irradiation facility of HANARO are described, and the major factors of the facility for BNCT applications such as the neutron flux and gamma-ray dose rate are determined.

## 2. Development of neutron irradiation facility

*2.1. Design considerations*

HANARO has seven horizontal beam tubes for neutron beam applications. The ex-core neutron irradiation facility for a BNCT has been installed by using a typical tangential beam tube whose nose is located in the thermal neutron peak area of the heavy water reflector tank. Because the beam tubes are narrow in width and long at over 400 cm, a sufficient amount of epithermal neutrons for a BNCT cannot be obtained. In order to obtain a sufficient thermal neutron field with a low level contamination of fast neutrons and gamma-rays, a radiation filtering method by using a single crystal, cooled by liquid nitrogen, has been adopted.

In HANARO, various experiments are undertaken simultaneously, and its high power density prevents a restart within about two days after a shutdown due to the xenon buildup. Therefore, an effective method to operate the irradiation facility without interference to the reactor operation has been developed.

In order to apply a local irradiation method after a debulking, temporary surgical operations before and after an irradiation for a patient should be possible at the irradiation site while the reactor is under a full power operation. Since there was no available space to perform an operation inside the HANARO reactor hall, an irradiation room has been designed so that a surgical operation might be possible in it. The irradiation room should be spacious enough for an operation, and the medical requirements must be met. Also, the radiation level outside the irradiation room should be maintained below the allowable level during an irradiation.

*2.2. Water shutter*

In order to perform the various research activities of HANARO without any interference caused by an operation of the BNCT facility, a water shutter has been designed and installed. The shutter was made of aluminium. Its inner diameter is 236 mm, and its axial length is 1350 mm. The volume of the inner part of the shutter is 60 liters. The water in the shutter is drained into the outer reservoir by compressed air, which is supplied by the electric pump. The time required for the water going into and out of the shutter is about 2-3 minutes. When the shutter is closed, the radiation dose rate inside the irradiation room is much lower than the allowable level even if the reactor is operated at a full power.

*2.3. Radiation filter*

*2.3.1. Filter design.* The purpose of the radiation filter is for a maximum extraction of the thermal neutron from the reactor with as small contamination of the fast neutrons and gamma-rays as possible. Even though fast neutrons and gamma-rays are also used for a cancer therapy, they cannot kill the cancer cell selectively like the BNCT, and thus, the fast neutron and gamma-ray extractions should be minimized.

Material used for this neutron filter should have a very small thermal neutron cross-section while its fast neutron cross-section is rather high. Table 1 shows several materials whose cross-sections for the fast neutrons are larger than those for the thermal neutrons (Nuclear Energy Agency 1994). From the table, it is confirmed that helium is the most advantageous from the aspect of a cross-section. However, helium is not proper for this application because it must be used in a gaseous phase. Aluminium was proposed as the next choice for the filter material, and thus we simulated the neutron transmission through the aluminium for the neutron beam extracted from a beam port of HANARO by using the MCNP code. However, the improvement of the flux ratio of the thermal neutron to the fast neutron was rather small.

Table 1. Several materials whose cross-sections for the fast neutrons are larger than those for the thermal neutrons.

| Material | Effective total cross-section [barns] | | Note |
| --- | --- | --- | --- |
|  | Maxwell averaged | Fission spectrum averaged |  |
| He-4 | 0.8490 | 3.660 | Gas |
| Al-27 | 1.640 | 3.380 | Abundance : 100% |
| Silicon | 2.256 | 3.414 |  |
| Sulfur | 1.470 | 2.677 |  |

The thermal neutron cross-section of a material with the crystal structure where the atoms are regularly arranged becomes smaller than that in the powder, while the change of the cross-section for the fast neutron is negligible. In addition, the penetration ability of thermal neutrons is enhanced when the single crystal is cooled to the temperature of liquid nitrogen ($LN_2$). In the above materials, the silicon single crystal is the only one that can be grown to a size large enough to be applicable to the filter. Therefore, a silicon single crystal cooled by liquid nitrogen was chosen as the neutron filter. The bismuth crystal was used after the silicon crystal as the gamma-ray filter. The bismuth crystal has a small neutron cross-section and a small secondary gamma-ray production rate. Figure 1 shows the neutron cross-sections of the silicon and the bismuth in the cases of room and liquid nitrogen temperatures. These cross-sections are deduced semi-empirically including the absorption and thermal diffusion scattering (Freund 1983). From the figure, it is confirmed that the shape of the cross-section of the silicon crystal is proper for the objective of this filter. The cross-sections of the bismuth shift to a lower energy than those of the silicon, and the rate of change is bigger than that of the silicon.

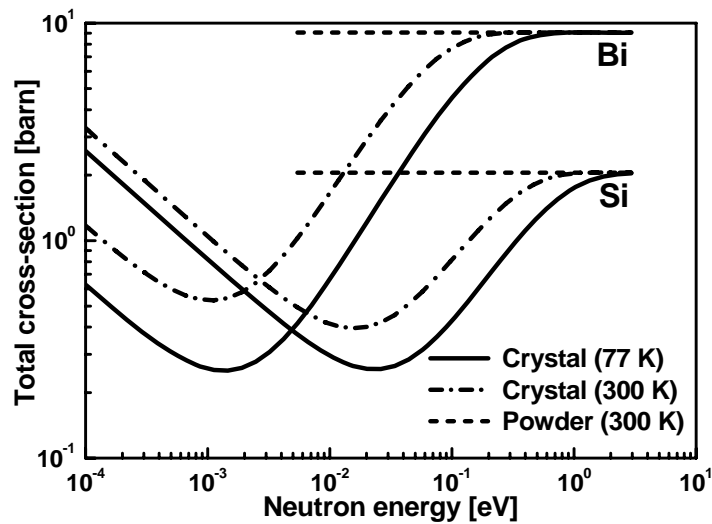

Fig. 1. Neutron cross-sections of the silicon and the bismuth in the cases of the room and liquid nitrogen temperatures.

After the sensitivity studies, the filter details were determined as a silicon single crystal of 40 cm in length and 20 cm in diameter and a bismuth single crystal of 15 cm in length and 10 cm in diameter. The silicon and bismuth crystals were fabricated so that the concentration of the dopants or impurities inside them was as low as possible. In order to cool down the crystals, a liquid nitrogen supply system and a vacuum system were installed. The vacuum gap for the thermal insulation was maintained below $1\times10^{-4}$ torr. The radiation filter was positioned at the exit part of the beam tube to minimize the neutron loss due to the direction change after scattering. The shielding surrounding the filter was composed of polyethylene, lead, plastics containing boron, and polycrystalline bismuth. Figure 2 shows the schematic layout of the radiation filter of the HANARO ex-core irradiation facility for the BNCT applications.

*2.3.2. Calculations for the beam parameters.* The thermal and fast neutron fluxes and the gamma-ray dose were calculated by using the MCNP code. The calculation model for the whole facility including the reactor core, the reflector tank, the beam tube and the radiation filter is too big and complicated to be dealt with by the MCNP. In order to obtain a reliable result with a small statistical deviation, the calculation was divided into two steps. At first, the neutron source at the nose of the beam tube and its angular distribution were determined by using a model of the reactor core and the reflector region. And then, the beam parameters were calculated through the beam tube by using a determined source at the nose. In the calculation for a long beam tube, a geometry splitting with Russian roulette was used to deliver more particles towards the beam exit for a variance reduction. The importance of each cell nearer to the beam exit was increased.

The cross-sections of the silicon and bismuth crystals including the case of a cooling to the liquid nitrogen temperature are not provided in the MCNP code. Therefore, we added their cross-section data, which were obtained by using a semi-empirical formula, to the MCNP library (Freund 1983).

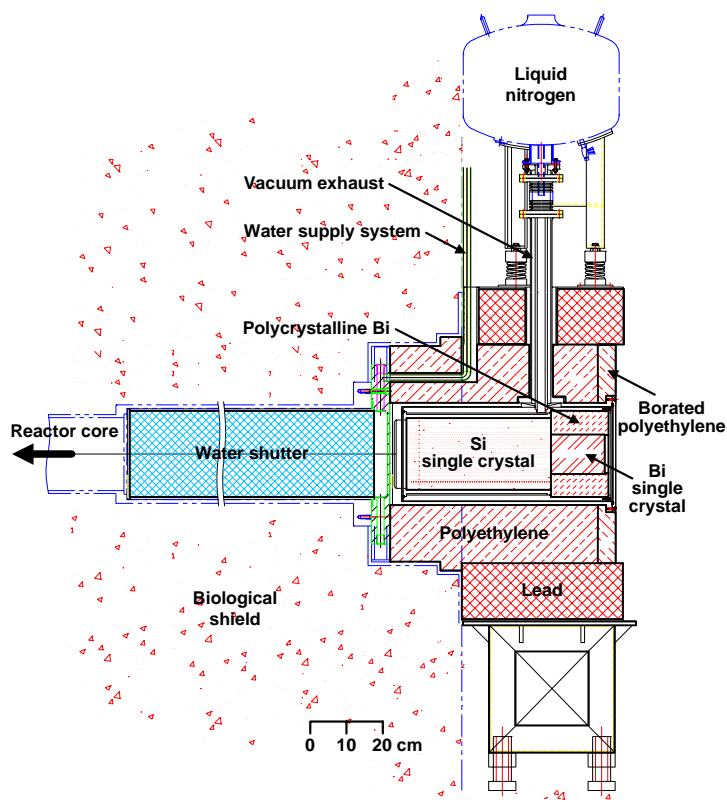

Fig. 2. Layout of the radiation filter of the HANARO ex-core neutron irradiation facility for the BNCT.

*2.4. Front shielding, collimator and beam monitoring system*
Since the radiation level inside the room should be low enough for the workers to access the irradiation room when the water shutter is closed, we installed the shielding structure in front of the radiation filter. Inside this shielding structure, the beam collimator and the detectors for monitoring the neutron beam and the gamma-rays were installed. When the water shutter is closed, the spatial radiation dose rate inside the irradiation room is lower than 5 μSv/hr.

The cone shaped collimators were manufactured by sintering a mixture of $^6Li_2CO_3$ powder (95 % enriched in Li-6) and high density polyethylene powder (Riley 2004). So far, we have made two collimators with beam hole diameters of 97 mm and 146.7 mm. The thickness of the sintered collimator is 15 mm or more. The diameter of the collimator used in the present experiments was 146.7 mm.

A fission chamber and an ion chamber were installed to monitor the changes of the neutron and gamma-ray extractions due to an unexpected reactor power variation or malfunctioning of the components of the facility such as the water shutter. The sizes of the detectors are 20.3 cm in length and 2.6 cm in diameter for the fission chamber and 4.4 cm in length and 1.9 cm in diameter for the ion chamber. Figure 3 represents the monitoring results for the extracted neutrons and gamma-rays. From the figure, it is confirmed that the two beam monitoring systems were operated successfully for a safe and exact irradiation for the whole power range of the reactor.

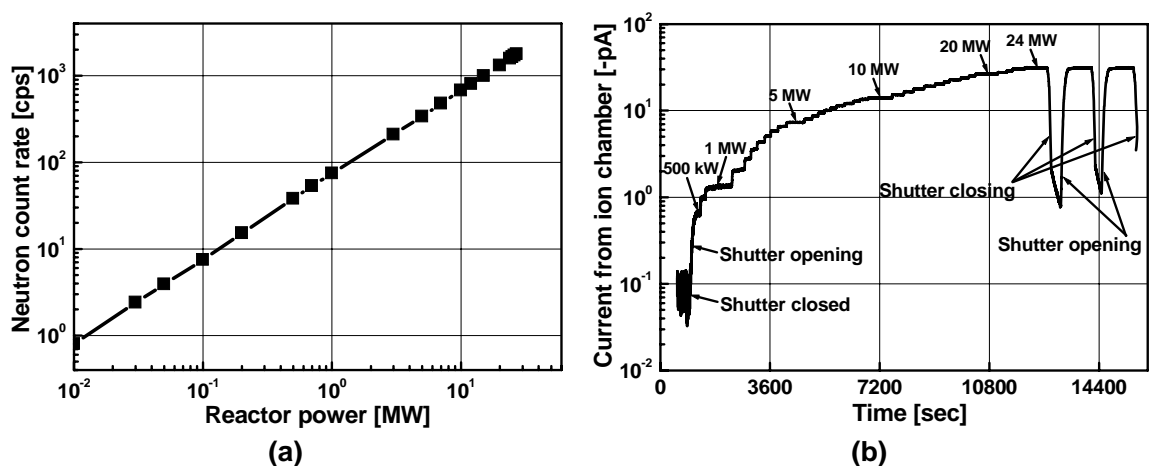

Fig. 3. Neutron count rate measured by the fission chamber with the reactor power(a) and the current from the ion chamber continuously monitoring the gamma-rays(b).

*2.5. Irradiation room*

The irradiation room was designed not only for the neutron irradiation but also for the temporary surgical operation. A sufficient space in front of the beam exit was prepared for various alignments of a patient's bed according to the irradiation position. A neutron shielding structure composed of borated plastic sheets and borosilicate glass will surround the patient during an irradiation. This neutron shielding tent prevents the materials inside the room from being activated, and it allows a medical team to access the room immediately after an irradiation. The patient can be monitored through the borosilicate glass via a video camera.

The radiation level in the room is kept sufficiently low during the preparation and the post actions of an irradiation due to the water shutter. During the beam extraction, the gamma-ray and neutron dose rates outside the irradiation room are also low enough for any other research activity of the reactor to be undertaken. The distance from the beam exit to the wall of the room is about 200 cm to 550 cm, and the height of the irradiation room is about 350 cm. Figure 4 shows an inside view of the irradiation room.

## 3. Measurements of the beam characteristics

*3.1. Neutron beam*

Experimental measurements of the neutron beam characteristics were performed by using bare and cadmium covered gold foils and wires (International Atomic Energy Agency 1970). First of all, the absolute neutron fluxes and the cadmium ratio were measured at several points on the surfaces of the beam exit hole, the Li-6 contained beam collimator and the front shielding by using gold foils. The thickness of the gold foil was 0.0254 mm, and its diameter was 12.7 mm. The cadmium ratio was measured by using the same gold foils and a cadmium cap with a thickness of 0.5 mm. Secondly, the radial neutron flux distribution at the exit surface was measured by using a bare gold wire and a gold wire inserted into a cadmium tube with a thickness of 0.5 mm. The diameters of the Au wires were either 0.1 or 0.254 mm. And then, the axial neutron flux distribution as a function of the axial distance from the beam exit surface was investigated. Finally, the two-dimensional in-phantom neutron flux distribution was measured. For a mapping of the neutron flux distribution inside the phantom, a number of gold monitors were located at appropriate positions inside the slab phantom. The phantom used in the experiments was a 300×300 mm$^2$ rectangular solid slab phantom. Its density was 1.045 g/cm$^3$, and it was composed of several slabs with thicknesses of 1, 2, 5 and 10 mm.

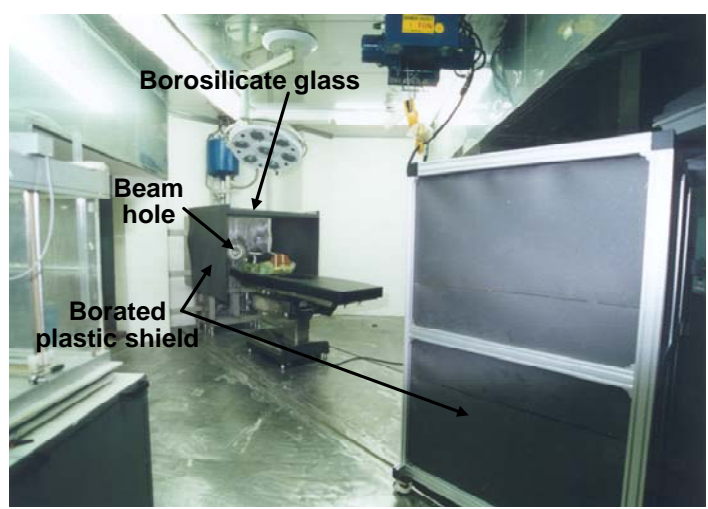

Fig. 4. Inside view of the irradiation room.

The time required to open the water shutter for a neutron irradiation was about 3 minutes, and that to close it was about 2 minutes. Thus, the irradiation time cannot be determined clearly, unlike a quick shutter system. In order to investigate the variation of the neutron irradiation rate at the flux monitor position, the variation of the count rate of the fission chamber was measured during a shutter operation. The neutron beam parameters were measured with the variation of the liquid nitrogen cooling condition of the radiation filter. All these experiments were performed at a reactor power of 24 MW, while the nominal power of HANARO is 30 MW.

The activity of the irradiated gold sample was measured by using a calibrated high purity germanium detector system. The uncertainty in determining the activity was within 2%, and the dominant factor was the uncertainty of the standard source activity used in the calibration procedure.

*3.2. Gamma-ray dose*
The gamma-ray dose of this facility was measured by using TLD-700 thermoluminescence dosimeter. The specifications of the TLD dosimeter used in this experiment are represented in table 2. TLD-700 dosimeter is usually assumed to measure the gamma-ray dose only because the content of Li-6 in it is below about 0.04%. However, the Li-6 induced signal within the TLD-700 may not be negligible when it is exposed to relatively large thermal neutron fluences like this BNCT irradiation (Attix 1986, Raaijmakers *et al* 1996).

In order to eliminate the thermal neutron contribution to the TLD-700 dosimeter, we irradiated the TLD-700 dosimeters covered with $^6$LiF tiles (Kumada *et al* 2004). The $^6$LiF tile (95 % enriched in Li-6) was made by a sintering procedure from the powder. The size of the tile was 4.0×4.0 cm$^2$, and its thickness was 4 mm. The density of the sintered tile was 2.25 g/cm$^3$, and the transmission of the neutron through the tile was negligible. The calibration of the TLD for the gamma-rays was performed by the Co-60 standard source. The dose values from the irradiated dosimeters were determined by using Harshaw Reader Model 3500.

Table 2. Specifications of the thermoluminescence dosimeter used in the measurements.

| Parameters | Figures |
|---|---|
| Type | TLD-700 |
| Materials | Lithium Fluoride (Li-7 isotope) LiF:Mg,Ti |
| Applications | Gamma, Beta |
| $Z_{eff}$ | 8.2 |
| TL emission spectra | 3500-6000 Angstrom |
| Sensitivity at Co-60 relative to LiF | 1.0 |
| Energy Response 30 keV/Co-60 | 1.25 |
| Useful Range | 10 μGy - 10 Gy |
| Fading | 5%/yr at 20 ℃ |
| Diameter | 1 mm |
| Length | 6 mm |

**4. Results**

*4.1. Characteristics of the neutron beam*

Figure 5 shows the relative count rate of the fission chamber (FC) installed inside the front shielding to monitor the beam extraction according to an operation of the water shutter. As shown in the figure, an additional irradiation during the shutter opening ($t_1 \sim t_2$) and closing ($t_3 \sim t_4$) should be considered to accurately estimate the total neutron fluence delivered to the target. The thermal neutron flux at this facility, $\phi_0$ should be deduced from the flux monitor which was irradiated with the shutter operation. We supposed that a neutron irradiation at a position of the flux monitor during the shutter operation varied linearly as shown by the solid line in figure 5. If $t_1$ is set to be zero, the neutron flux can be deduced from the activity of the flux monitor as follows.

$$\phi_0 = \frac{\lambda}{N_m \sigma} \frac{A(t_4)}{\left[ \frac{1}{t_2}(e^{-\lambda t_4} - e^{\lambda(t_2 - t_4)}) + \frac{1}{t_3 - t_4}(e^{\lambda(t_3 - t_4)} - 1) \right]}, \quad (1)$$

where, $A$ is the activity of the flux monitor, $N_m$ is the number of the target nucleus at the flux monitor, $\sigma$ is the radiative capture cross-section, and $\lambda$ is the decay constant of the reaction product. In these experiments, we used gold foils and wires as flux monitors.

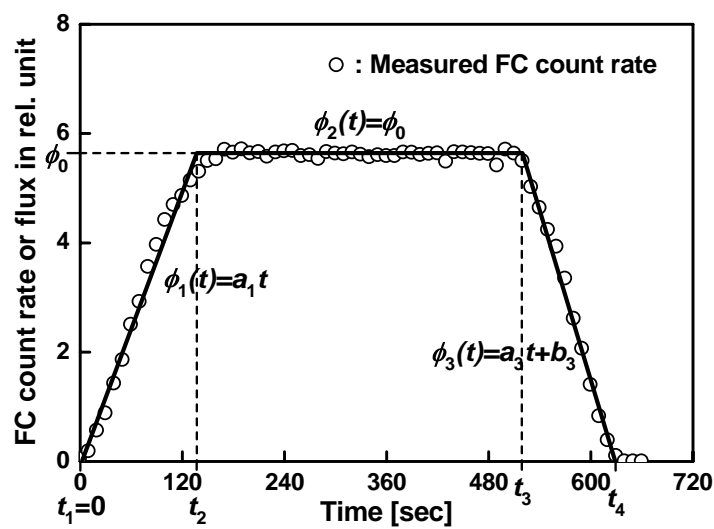

Fig. 5. Relative count rate of the fission chamber (FC) according to the operation of the water shutter.

Since a mono-directional neutron beam is extracted through the long beam tube, the neutron attenuation effect inside the gold flux monitors was calculated by considering the pathlength of the neutron inside the monitor. The correction factors for the neutron attenuation were 0.9758 for the gold wire with a 0.1 mm in diameter and 0.9402 for that with a 0.254 mm in diameter, respectively. In the case of the gold foil with a thickness of 0.0254 mm, the correction was negligible.

Table 3 shows the thermal neutron flux measured by using gold foils at several positions on the surface of the beam exit hole and the surroundings. The radiation filter was at room temperature, and the reactor power was 24 MW. The neutron flux outside the beam collimator is much smaller than that inside the beam extraction area. The cadmium ratio measured at the center of the beam exit hole is 104.

Table 3. Neutron flux measured by using gold foils at the beam exit.

| Position | Radial distance from center of beam exit [cm] | Measured thermal neutron flux [n/cm$^2$s] |
|---|---|---|
| Beam exit hole | 0 | $(8.34\pm0.21)\times10^8$ |
|  | 2.5 | $(7.99\pm0.20)\times10^8$ |
|  | 5.5 | $(7.03\pm0.18)\times10^8$ |
| Collimator | 8.0 | $(1.43\pm0.038)\times10^7$ |
| Shielding | 10.1 | $(1.28\pm0.035)\times10^6$ |
|  | 19.0 | $(6.06\pm0.17)\times10^5$ |
|  | 29.0 | $(3.63\pm0.10)\times10^5$ |
|  | 39.0 | $(2.38\pm0.072)\times10^5$ |

Figure 6 shows the relative activity distributions of bare and cadmium covered gold wires irradiated at the beam exit surface in order to measure the neutron beam distributions. The activity distributions of the wires were scanned by a HPGe detector with the lead shielding and a front slit of 4 mm in width. The area of the 411 keV gamma-ray peak from the Au-198 was more than 3,000. As shown in the figure, it is confirmed that both the thermal and fast neutron irradiations outside the collimation area are negligible. The neutron irradiation is concentrated well inside the aperture of the Li-6 contained beam collimator. In the case of the collimator of 15 cm in diameter, the radial neutron beam distributions inside ±5 cm from the beam center seem to be flat. In that region, the maximum deviation between the measurements is 7%. The uncertainty of the activity measurement is about 3~4%, including the uncertainties arising from the non-uniformity in the gold wire diameter and a possible bending of the wire.

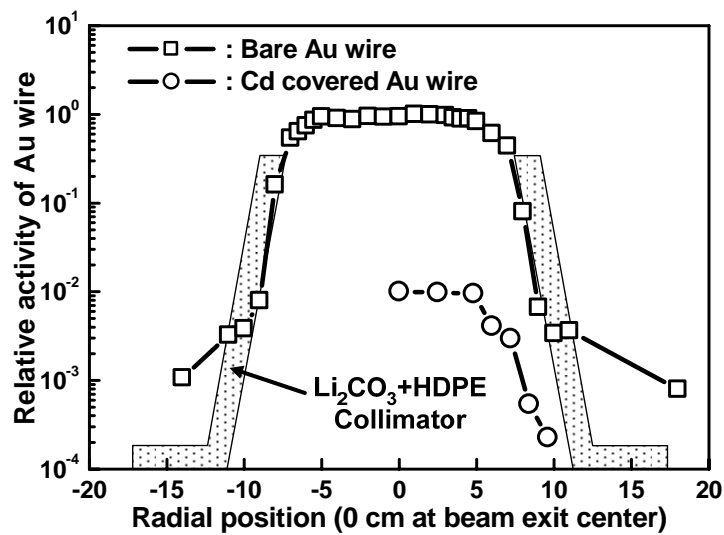

Fig. 6. Relative activity distributions of bare and cadmium covered gold wires irradiated at the beam exit surface.

The absolute neutron flux and Cd ratio measured at different experimental conditions are represented in table 4. The uncertainty of the measured flux is within 2.5%. When the radiation filter is cooled with liquid nitrogen, the neutron flux and Cd ratio increase by over 40%. It means that the increase of the thermal neutron flux due to the cooling of the radiation filter is rather big. From figure 3(a), we expect that the neutron flux from the beam exit would be linearly proportional to the reactor power. Thus, we can estimate the thermal neutron fluxes at the reactor power of 30 MW from the measurements at 24 MW, and they are represented in table 4. For the free beam extraction without phantom, the differences between the expected values from the measurements and the calculated ones by using the MCNP are less than 10% regardless of the cooling condition of the radiation filter. These differences are very small when considering the various uncertainty sources in the MCNP calculation and the experimental condition. In the experiment, the main contributions to this difference are the uncertainty in the temperature of the crystal of the radiation filter and the change of the reactor core conditions such as a disposition of the nuclear fuel. In the calculation, the neutron cross-section of the bismuth single crystal may have a big uncertainty because it has high density of reciprocal lattice points and the large mosaic spread (Rustad *et al* 1965). It is concluded that the thermal neutron flux and the Cd ratio which can be obtained at this facility are $1.49 \times 10^9$ n/cm$^2$s and 152, respectively, at 30 MW reactor power.

Table 4. Absolute neutron flux and Cd ratio at several experimental conditions.

| Experimental conditions | Measured thermal neutron flux at 24 MW [n/cm$^2$s] | Measured Cd ratio | Thermal neutron flux at 30 MW [n/cm$^2$s] | |
|---|---|---|---|---|
| | | | Expected from measurements | Calculated by using MCNP |
| No LN$_2$ cooling, no phantom | $0.83 \times 10^9$ | 104 | $1.04 \times 10^9$ | $0.94 \times 10^9$ |
| With LN$_2$ cooling, no phantom | $1.19 \times 10^9$ | 152 | $1.49 \times 10^9$ | $1.48 \times 10^9$ |
| No LN$_2$ cooling, with phantom | $1.74 \times 10^9$ | | $2.18 \times 10^9$ | |
| With LN$_2$ cooling, with phantom | $2.23 \times 10^9$ | | $2.79 \times 10^9$ | $3.67 \times 10^9$ |

When the phantom is installed in front of the beam exit, the measured neutron flux at the phantom surface increases by twofold or more than that without phantom. The calculated thermal neutron flux on the phantom surface is 30% larger than the measured value. The difference may be attributed to the uncertainties of the physical and nuclear data of the phantom materials, the characteristics of the MCNP calculation, the self shielding effect of the gold samples and the different core conditions. The uncertainty of the MCNP result at the boundary of the medium materials with very different nuclear characteristics may be increased due to its tally averaging characteristics. The self shielding effects of the gold foil and wire irradiated under a parallel neutron beam can be deduced easily. When a phantom is installed, the number of scattered neutrons by the phantom is increased considerably and the directional changes of the scattered neutrons will make their pathlengths in the sample longer. Further study is required to examine the cause of the difference between the measurements and the calculation through a precise analysis.

Figure 7 shows the measured and calculated neutron fluxes as a function of the distance from the beam exit surface when the radiation filter is maintained at room temperature, and the reactor power is 24 MW. In the figure, the solid line represents the fitting result to the calculated values over a distance of 45 cm from the beam exit with the following equation.

$$y = \frac{a}{(x+458)^b} + c \quad (b=2), \tag{2}$$

where, 458 means the distance from the nose to the exit of the beam tube in cm. This fitting line is not consistent with the calculated values within 45 cm. The main contribution to the neutron flux over 45 cm is the neutrons that are extracted undergoing very small angle collisions or without any collision through the beam path from the nose. Therefore, the neutron flux over 45 cm follows the $1/r^2$ law, where $r$ is the distance from the source. Near to the beam exit, the neutrons scattered by the surrounding materials such as silicon, bismuth and collimator are added to the $1/r^2$ component of the neutron flux. Therefore, the neutron flux abruptly increases when approaching the beam exit from outside.

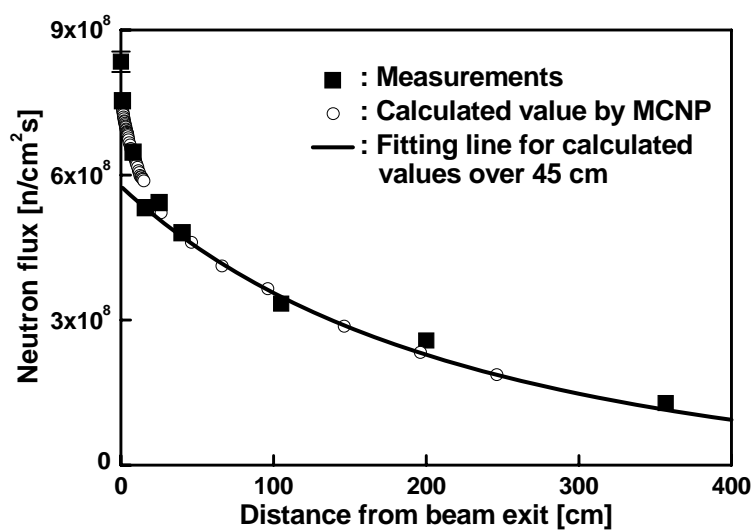

Fig. 7. Measured and calculated neutron fluxes as a function of the distance from the beam exit surface.

Figure 8 shows the measured and the calculated neutron flux distributions as a function of the depth in the phantom along the central axis of the beam exit. The measured thermal neutron flux reaches a maximum near a 3 mm depth in the phantom, and then it decreases rapidly. The maximum flux is about 25 % larger than that at the phantom surface, and the measured value at a depth of 22 mm in the phantom is about a half of the maximum value. When the radiation filter is cooled by liquid nitrogen ($LN_2$), the thermal neutron flux is 30% larger than that without cooling. As shown in the table 4 and figure 8, it is confirmed that the calculated neutron fluxes at the surface or inside the phantom are about 30% larger than the measured ones while the two neutron flux distributions show the same trend.

The measured two-dimensional in-phantom neutron flux distribution is shown in figure 9. Significant neutron irradiation is observed within 20 mm from the phantom surface (Yamamoto *et al* 2002, Kim *et al* 2004). The area enclosed by the isoflux line of $2\times10^9$ n/cm$^2$s or more is located within a 12 mm depth in the phantom. The $2\times10^9$ n/cm$^2$s flux is about 77 % relative to the maximum flux. The neutron flux distribution near the phantom surface looks as though it has a double isoflux area enclosed by the $2.6\times10^9$ line in the lateral direction. This area has the maximum neutron flux, and it is caused by the flux peaking inside the phantom due to the scattered neutrons by the collimator. The thermal neutron flux falls off rapidly outside 50 mm in the lateral direction. The measured neutron flux distribution inside the phantom may be utilized in the irradiation planning, such as determining the optimal size of the collimator and the optimal position for a patient's head.

*4.2. Gamma-ray dose*
The gamma-ray dose at this facility is measured with TLD-700 dosimeter by eliminating the effect of the thermal neutron by using $^6$LiF tile. The contribution of the fast neutron is neglected due to its low flux and small reaction cross-sections. Table 5 shows the measured gamma-ray dose rates at several points on the beam exit surface when the radiation filter is cooled by liquid nitrogen. Since the reactor was operated at 24 MW, the gamma-ray dose rate at 30 MW, the nominal power of HANARO, is expected to be about 80 cGy/hr. The gamma-ray dose rate outside of the beam collimator is negligible.

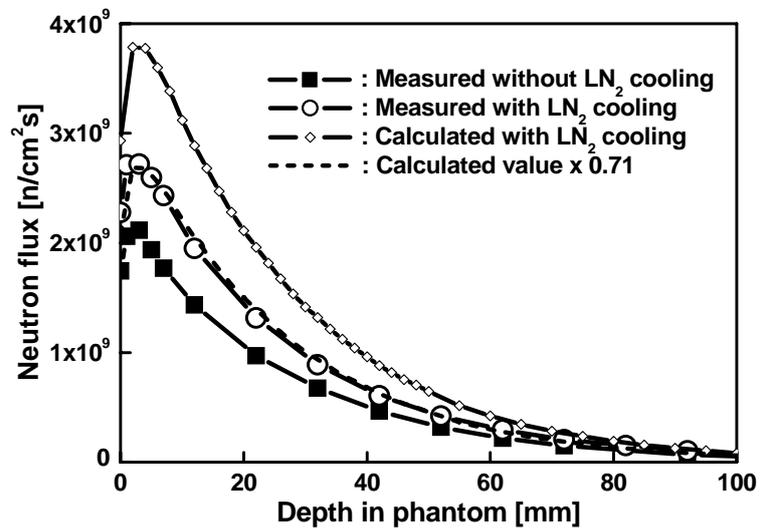

Fig. 8. Measured and calculated neutron flux distributions as a function of the depth in the phantom.

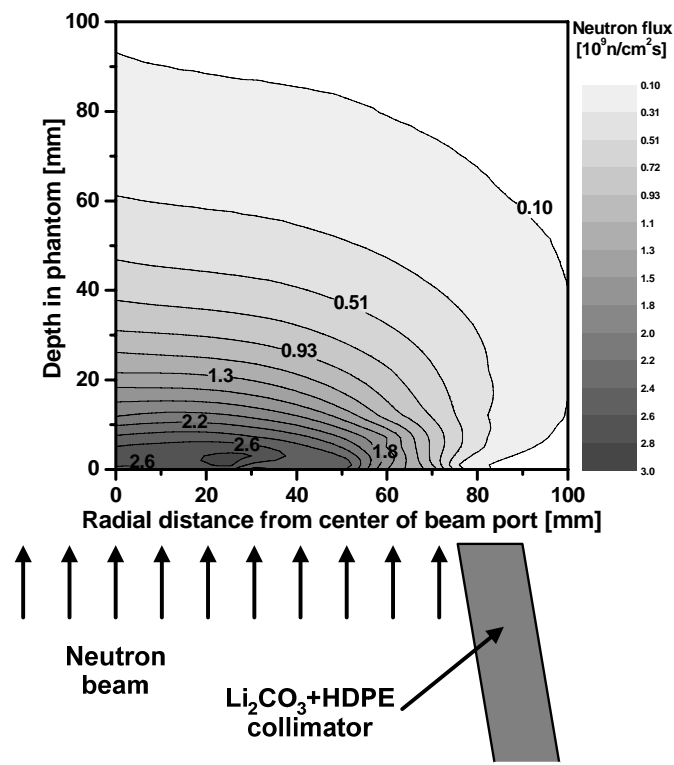

Fig. 9. Measured two-dimensional in-phantom neutron flux distribution.

Figure 10 shows the gamma-ray dose rate distributions extracted from this facility as a function of the distance from the beam exit. The measurements agree well with the prediction by the MCNP calculation. The fractional standard deviations of the calculated values at 26 cm and 46 cm from the beam exit are 6.8% and 12.2% respectively, and those inside 15 cm are less than 4.8%.

Table 5. Determined gamma-ray dose rate at several points on the beam exit surface.

| Position | Radial distance from center of beam exit [cm] | Gamma-ray dose rate [cGy/hr] |
| --- | --- | --- |
| Beam exit | 0 | 65.4±3.38 |
| Collimator | 7.5 | 15.0±1.18 |
| Shielding | 10.0 | 2.60±0.235 |

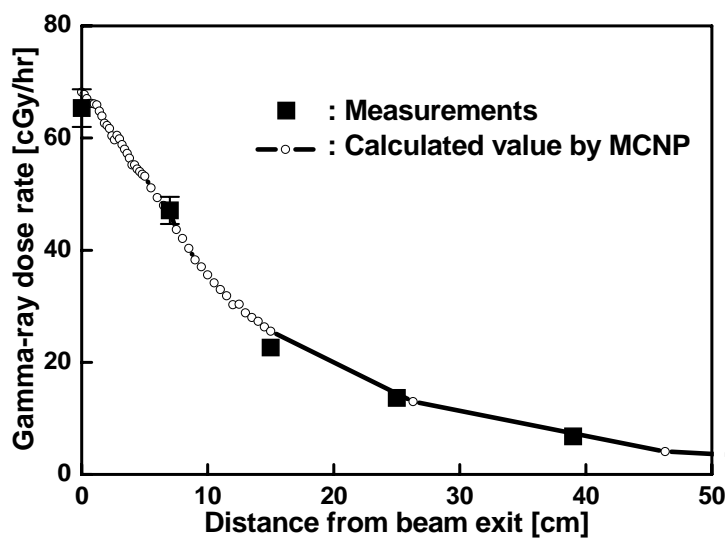

Fig. 10. Measured and calculated gamma-ray dose rate distributions as a function of the distance from the beam exit.

## 5. Conclusion

The HANARO ex-core neutron irradiation facility was developed by using a tangential beam tube for various applications in the boron neutron capture therapy (BNCT) field. Considering that not enough epithermal neutrons for the BNCT can be obtained at the irradiation position due to the design characteristics of HANARO, a thermal neutron facility for various applications was designed by means of a fast neutron and gamma-ray filtering method. The radiation filter consists of a silicon single crystal for a maximum extraction of a thermal neutron and a bismuth crystal for a gamma-ray shielding. A water shutter in front of the radiation filter was installed in order to perform various research activities without any interference due to an operation of the BNCT facility. The basic facilities for the BNCT were installed and the irradiation room was constructed where even a temporal surgical operation could be conducted.

The major parameters for the BNCT application such as the neutron flux distributions were determined. The thermal neutron flux and the Cd ratio available from this facility for a condition of no phantom and the radiation filter cooled by liquid nitrogen are $1.49 \times 10^9$ n/cm$^2$s and 152, respectively. The gamma-ray dose rate at the beam exit for a free beam condition is about 80 cGy/hr. This HANARO thermal neutron facility can be used not only for a clinical trial but also for various treatment researches of the BNCT field.

**Acknowledgments**

The authors acknowledge the reading of exposed TLD of Mr. D.H. Lee and Ms. S.H. Seo at Korea Institute of Radiological and Medical Sciences. This work was supported by the Ministry of Science and Technology, Republic of Korea.

**References**

Attix F H 1986 Introduction to radiological physics and radiation dosimetry (New York: John Wiley & Sons) p.47

Bisceglie E, Colangelo P, Colonna N, Santorelli P and Variale V 2000 On the optimal energy of epithermal neutron beams for BNCT *Phys. Med. Biol.* **45** 49-58

Byun S H, Sun G M and Choi H D 2002 Development of a prompt gamma activation analysis facility using diffracted polychromatic neutron beam *Nucl. Instr. and Meth.* **A487** 521-529

Freund A K 1983 Cross-sections of materials used as neutron monochromators and filters *Nucl. Intstr. and Meth.* **213** 495-501

International Atomic Energy Agency 1970 Neutron fluence measurements *Technical Reports Series No. 107* (Vienna: IAEA)

Kim M S, Park S J and Jun B J 2004 Measurements of in-phantom neutron flux distribution at the HANARO BNCT facility *J. Kor. Nucl. Soc.* **36** 203-209

Kumada H, Yamamoto K, Matsumura A, Yamamoto T, Nakagawa Y, Nakai K and Kageji T 2004 Verification of the computational dosimetry system in JAERI (JCDS) for boron neutron capture therapy *Phys. Med. Biol.* **49** 3353-3365

Nuclear Energy Agency 1994 Table of simple integral neutron cross section data from JEF-2.2, ENDF/B-VI, JENDL-3.2, BROND-2 and CENDL-2 *JEF Report 14* (Paris: OECD)

Raaijmakers C P J, Watkins P R D, Nottelman E L, Verhagen H W, Jansen J T M, Zoetelief J and Mijnheer B J 1996 The neutron sensitivity of dosimeters applied to boron neutron capture therapy *Med. Phys.*, **23(9)** 1581-1589

Riley K J, Binns P J, Ali S J and Harling O K 2004 The design, construction and performance of a variable collimator for epithermal neutron capture therapy beam *Phys. Med. Biol.* **49** 2015-2028

Rustad B M, Als-Nielsen J, Bahnsen A, Christensen C J and Nielsen A 1965 Single-crystal filters for attenuating epithermal neutrons and gamma rays in reactor beams *Rev. Sci. Intstr.* **36** 48-54

Sakurai Y and Kobayashi T 2000 Characteristics of the KUR Heavy Water Neutron Irradiation Facility as a neutron irradiation field with variable energy spectra *Nucl. Instr. and Meth.* **A453** 569-596

Sauerwein W, Moss R and Wittig A ed 2002 *Research and Development in Neutron Capture Therapy* (Bologna: Monduzzi)

Slatkin D N 1991 A history of boron neutron capture therapy of brain tumors Brain 114 1609

Yamamoto T, Matsumura A, Yamamoto K, Kumada H, Shibata Y and Nose T 2002 In-phantom two-dimensional thermal neutron distribution for intraoperative boron neutron capture therapy of brain tumours *Phys. Med. Biol.* **47** 2387-2396

Zamenhof R G, Coderre J A, Rivard M J and Patel H 2004 Eleventh world congress on neutron capture therapy *Appl. Radiat. Isot.* **61** 731